\documentclass[twocolumn,tighten,times]{aastex62}
\usepackage{mathrsfs}
\usepackage{amsmath}
\usepackage{url}
\usepackage[normalem]{ulem}
\usepackage{graphicx}
\usepackage{float}

\usepackage{natbib}
\usepackage{color}
\bibpunct{(}{)}{;}{a}{}{,}

\newcommand\aproxgt{\mathrel{%
      \rlap{\raise 0.511ex \hbox{$>$}}{\lower 0.511ex \hbox{$\sim$}}}}
\newcommand\aproxlt{\mathrel{%
      \rlap{\raise 0.511ex \hbox{$<$}}{\lower 0.511ex \hbox{$\sim$}}}}

\newcommand{\ignore}[1]{}
\newcommand{\storm}{{STORM}}
\newcommand{\swift}{{\it Swift}}
\newcommand{\hst}{{\it HST\,}}
\newcommand{\xmm}{{\it XMM-Newton\,}}
\newcommand{\nustar}{{\it NuSTAR\,}}

\newcommand{\ngc}{NGC\,5548}
\newcommand{\et}{{et~al.\, }}

\newcommand{\civ}{\mbox{\rm C\,{\sc iv}}}

\newcommand{\heii}{\mbox{\rm He\,{\sc ii}}}

\begin{document}

\title{Space Telescope and Optical Reverberation Mapping Project. XI.  Disk-wind characteristics and  contributions to the very broad emission lines of NGC 5548}

\author[0000-0002-0964-7500]{M.~Dehghanian}
\affiliation{\ignore{UKy}Department of Physics and Astronomy, The University of Kentucky, Lexington, KY 40506, USA}

\author[0000-0003-4503-6333]{G.~J.~Ferland}
\affiliation{\ignore{UKy}Department of Physics and Astronomy, The University of Kentucky, Lexington, KY 40506, USA}

\author[0000-0002-2180-8266]{G.~A.~Kriss}
\affiliation{Space Telescope Science Institute, 3700 San Martin Drive, Baltimore, MD 21218, USA}

\author[0000-0001-6481-5397]{B.~M.~Peterson}
\affiliation{Space Telescope Science Institute, 3700 San Martin Drive, Baltimore, MD 21218, USA}
\affiliation{Department of Astronomy, The Ohio State University, 140 W 18th Ave, Columbus, OH 43210, USA}
\affiliation{Center for Cosmology and AstroParticle Physics, The Ohio State University, 191 West Woodruff Ave, Columbus, OH 43210, USA}

\author[0000-0003-0944-1008]{K.~T.~Korista}
\affiliation{\ignore{WM}Department of Physics, Western Michigan University, 1120 Everett Tower, Kalamazoo, MI 49008-5252, USA}

\author[0000-0002-2908-7360]{M.~R.~Goad}
\affiliation{\ignore{Leicester}Department of Physics and Astronomy, University of Leicester,  University Road, Leicester, LE1 7RH, UK}

\author[0000-0002-8823-0606]{M.~Chatzikos}
\affiliation{\ignore{UKy}Department of Physics and Astronomy, The University of Kentucky, Lexington, KY 40506, USA}

\author[0000-0002-2915-3612]{F.~Guzm\'{a}n}
\affiliation{\ignore{UKy}Department of Physics and Astronomy, The University of Kentucky, Lexington, KY 40506, USA}

\author[0000-0003-3242-7052]{G.~De~Rosa}
\affiliation{Space Telescope Science Institute, 3700 San Martin Drive, Baltimore, MD 21218, USA}

\author[0000-0002-4992-4664]{M.~Mehdipour}
\affiliation{\ignore{SRON}SRON Netherlands Institute for Space Research, Sorbonnelaan 2, 3584, CA Utrecht, The Netherlands}

\author[0000-0001-5540-2822]{J.~Kaastra}
\affiliation{\ignore{SRON}SRON Netherlands Institute for Space Research, Sorbonnelaan 2, 3584 CA Utrecht, The Netherlands}
\affiliation{\ignore{Leiden}Leiden Observatory, Leiden University, PO Box 9513, 2300 RA Leiden, The Netherlands}

\author{S.~Mathur}
\affiliation{Department of Astronomy, The Ohio State University, 140 W 18th Ave, Columbus, OH 43210, USA}
\affiliation{Center for Cosmology and AstroParticle Physics, The Ohio State University, 191 West Woodruff Ave, Columbus, OH 43210, USA}

\author[0000-0001-9191-9837]{M.~Vestergaard}
\affiliation{\ignore{Dark}DARK, Niels Bohr Institute, University of Copenhagen, Vibenshuset, Lyngbyvej 2, DK-2100 Copenhagen \O, Denmark}
\affiliation{\ignore{Steward}Steward Observatory, University of Arizona, 933 North Cherry Avenue, Tucson, AZ 85721, USA}

\author[0000-0002-6336-5125] {D. Proga}
\affiliation{Department of Physics \& Astronomy, University of Nevada, Las Vegas, 4505 S. Maryland Pkwy, Las Vegas, NV, 89154-4002, USA}

\author {T. Waters}
\affiliation{Department of Physics \& Astronomy, University of Nevada, Las Vegas, 4505 S. Maryland Pkwy, Las Vegas, NV, 89154-4002, USA}

\author[0000-0002-2816-5398]{{M.~C.~Bentz}}
\affiliation{\ignore{Georgia}Department of Physics and Astronomy, Georgia State University, 25 Park Place, Suite 605, Atlanta, GA 30303, USA}

\author[0000-0003-3746-4565]{S.~Bisogni}
\affiliation{INAF - Istituto di Astrofisica Spaziale e Fisica Cosmica Milano, via Corti 12, 20133 Milano, Italy }

\author[0000-0002-0167-2453]{W.~N.~Brandt}
\affiliation{\ignore{Eberly}Department of Astronomy and Astrophysics, Eberly College of Science, The Pennsylvania State University, 525 Davey Laboratory, University Park, PA 16802, USA}
\affiliation{Department of Physics, The Pennsylvania State University, 104 Davey Laboratory, University Park, PA 16802, USA}
\affiliation{\ignore{IGC}Institute for Gravitation and the Cosmos, The Pennsylvania State University, University Park, PA 16802, USA}

\author[0000-0001-9931-8681]{E.~Dalla~Bont\`{a}}
\affiliation{\ignore{Padova}Dipartimento di Fisica e Astronomia ``G. Galilei,'' Universit\`{a} di Padova, Vicolo dell'Osservatorio 3, I-35122 Padova, Italy}
\affiliation{\ignore{INAF}INAF-Osservatorio Astronomico di Padova, Vicolo dell'Osservatorio 5 I-35122, Padova, Italy}

\author[0000-0002-9113-7162]{M. M. Fausnaugh}
\affiliation{Department
of Physics, and Kavli Institute for Astrophysics and Space Research, Massachusetts Institute of Technology, Cambridge, MA 02139, USA}

\author[0000-0001-9092-8619]{J.~M.~Gelbord}
\affiliation{Spectral Sciences Inc., 4 Fourth Ave., Burlington, MA 01803, USA}

\author[0000-0003-1728-0304]{Keith~Horne}
\affiliation{\ignore{SUPA}SUPA Physics and Astronomy, University of St. Andrews, Fife, KY16 9SS Scotland, UK}

%\author{\red{C.~Knigge}}
%\affiliation{\ignore{Southampton}School of Physics and Astronomy, University of Southampton, Highfield, Southampton, SO17 1BJ, UK}

\author{I.~M.~M$^{\rm c}$Hardy}
\affiliation{\ignore{Southampton}School of Physics and Astronomy, University of Southampton, Highfield, Southampton, SO17 1BJ, UK}

\author[0000-0003-1435-3053]{R.~W.~Pogge}
\affiliation{Department of Astronomy, The Ohio State University, 140 W 18th Ave, Columbus, OH 43210, USA}
\affiliation{Center for Cosmology and AstroParticle Physics, The Ohio State University, 191 West Woodruff Ave, Columbus, OH 43210, USA}

\author{D.~A.~Starkey}
\affiliation{\ignore{SUPA}SUPA Physics and Astronomy, University of St. Andrews, Fife, KY16 9SS Scotland, UK}
%\affiliation{\ignore{Illinois}Department of Astronomy, University of Illinois Urbana-Champaign, 1002 W. Green Street, Urbana, IL 61801, USA}

\shorttitle{AGN \storm\ ~XI. The disk wind in \ngc}
\shortauthors{Dehghanian \et}

\begin{abstract}
In 2014 the NGC 5548  Space Telescope and Optical 
Reverberation Mapping campaign
discovered a two-month
anomaly when variations in
the absorption and
emission lines decorrelated
from continuum variations.
During this time the soft
X-ray part of the 
intrinsic spectrum had been
strongly absorbed by
a line-of-sight (LOS)
obscurer, which was
 interpreted as the upper
part of a disk wind 
. 

Our first paper showed that changes in
the LOS obscurer produces
the decorrelation between
the 
absorption lines and the 
continuum. 
A second study showed that
the base 
of the wind shields the BLR, 
leading to the emission-line decorrelation 
. 
In that study, we proposed the wind 
is normally transparent with no effect on the
spectrum. 
Changes in the wind
properties alter its
shielding and
affect the  SED  striking the BLR,
producing the observed
decorrelations. 

In this work we investigate the  impact of a translucent 
wind on the emission lines.
We simulate the obscuration  using \xmm, \nustar, and \hst\ observations to determine the physical characteristics of the wind.
We find that a translucent wind can contribute a part
of the He II and Fe K$\alpha$ emission.
It has a modest optical depth to electron scattering, which
explains the fainter far-side emission in the 
observed velocity delay maps.
The wind produces the very broad base seen in the UV emission lines
and may also be present in the Fe K$\alpha$ line. Our results highlight the importance of accounting for the effects of such  winds in the analysis of the physics of the central engine.

\end{abstract}


\section{INTRODUCTION } 

The broad emission-line
region (BLR) is closely
associated with the central
regions and the 
supermassive black hole
(SMBH) in AGNs.
Reverberation mapping
(RM) \citep{BMcK82,Pet93}
can determine the geometry and
kinematics of the BLR, which
can be used to infer the
mass of the BH
\citep{Horn04}. RM uses 
the time delay between the
continuum and emission-line
variations to determine 
the responsivity-weighted distance to the
line emitting region \citep{Pet04}, which is
commonly taken to represent a characteristic size
scale of the BLR. The time delay is, in fact, the
travel time of the ionizing photons from the
inner accretion disk region to the BLR gas.
The duration of the delay depends on the causal connection between the broad emission line gas and the ionizing continuum emission. This causal connection is one of the fundamental principles of RM.

In 2014, the most intensive
RM campaign, AGN STORM
(Space Telescope and Optical
Reverberation
Mapping; \citealt{DeRosa15,
Edelson15,Fau16, Goad16,
Pei17, Starkey17, Mathur17,
Kriss19, Deh19a}),
observed the AGN NGC 5548
for six months. This unique
dataset has revealed 
several unexpected results. 
For a period of
$\sim 2$ months mid-way through the
campaign, the
continuum and broad
emission-line variations
were observed to decorrelate
\citep{Goad16}, the so-called ``
emission-line holiday". At
almost the same time, the
continuum and narrow
absorption lines also
decorrelated
\citep{Kriss19}, the ``absorption-line holiday".
These spectral holidays,
along with the presence of 
an X-ray obscurer in our
line of sight (LOS) to the
SMBH \ \citep{Kaastra14},
distinguish the 2014 version
of NGC 5548 from normal AGNs.
There is no part of the
standard AGN scenario that
produces holidays, so 
clearly something
fundamental is missing
\citep[hereafter D19a \&
D19b]{Deh19a,Deh19b}.  This
is an 
opportunity to determine the
physics controlling the spectral holiday, to study AGN feedback and 
develop scenarios about 
this central activity that
affects the evolution of
galaxies. 

D19a show that the
variation of the LOS
obscurer covering factor
(CF) produces the observed 
absorption-line holiday.
\swift \  observations
\citep{Mehd16} show
that the absorption line
variations correlate with
the CF (figure 12 of D19a), so are consistent with this interpretation. D19b propose that 
the LOS obscurer is the
upper part of a symmetric
cylindrical disk-wind that
originates
from the inner parts of the
accretion disk and is
interior to the BLR.  As
argued by D19b, the base of
the 
wind forms an equatorial
obscurer, filtering the SED
before the ionizing photons strike the BLR, 
leading to the observed emission-line
holiday.

In this work, we
create potential models for the
equatorial obscurer. 
Unlike the LOS obscurer,
which can be studied by its
absorption of the SED, 
the geometry and
characteristics of the base
of the wind are unknown. 
It does not absorb along our
LOS, however, it filters the
SED of the photons that reach the BLR. In the
following Section we
use STORM BLR observations 
to infer the properties of
the obscurer.  
We use these constraints to
narrow down the parameters
and we propose a final model that not
only 
reproduces the emission-line
holiday (D19b), but is also
consistent with the 
observations, 
while 
reproducing the absorption-line
holiday (D19a).

Our preferred model of the base
of the wind is also a major
contributor to the observed
broad iron K$\alpha$ line.
Both disk winds 
and broad Fe K$\alpha$
emission are considered to be common
properties of AGNs, and we
propose that the SED
filtering through the wind
is too.

\section{Physical models of
the equatorial obscurer}
In this paper we consider
new models of the equatorial
obscurer.  We do not provide
new models of the BLR but
rather rely on the results
of D19b. Figure 4 of D19b 
shows that the equatorial
obscurer will lead to a
holiday if hydrogen is
fully ionized and a He$^+$
ionization front is present
within it (their
Case 2). All models in this
paper have a column density
adjusted so 
that the optical depth is 8
at 4 Rydberg. This optical
depth belongs to the left
threshold of Case 2 in D19b,
and ensures the presence of the
emission-line holiday. 

We adopt the SED of
\cite{Mehd15} in {\tt
CLOUDY} \citep[developer
version,][]{Ferland17}
and an open
geometry\footnote{Refer to section
2.3.4 of the {\tt CLOUDY}’s
documentation, 
 \citep{Ferland17}} 
for the equatorial 
obscurer.  An open geometry
is appropriate when the
emission-line cloud 
CF is small since diffuse
emission is assumed to
escape from the AGN.  The 
global BLR covering factor
is about 50$\%$ 
\citep[integrated
cloud covering fraction,
][]{Korista00} and the
equatorial obscurer must 
cover at least this much. 
So, it is intermediate
between an open and 
closed geometry. Inspired by
figure 1 of D19b, we adopt
an open geometry. In order
to make our predictions more 
accurate, we increased the
number of levels to $n=100$
for H like atoms. 
This allows a better
representation of the
collision physics that
occurs
within higher levels of the
atom. We also set the
spectral resolution to 5000
km s$^{-1}$. Changing the 
velocity width does not
resolve the lines but 
changes the
line-to-continuum contrast
ratio to simulate a
spectrometer 
measuring an unresolved
line. We further assume
photospheric solar
abundances
\citep{Ferland17}.  

\begin{figure*}
\centering
\includegraphics [width=\textwidth]{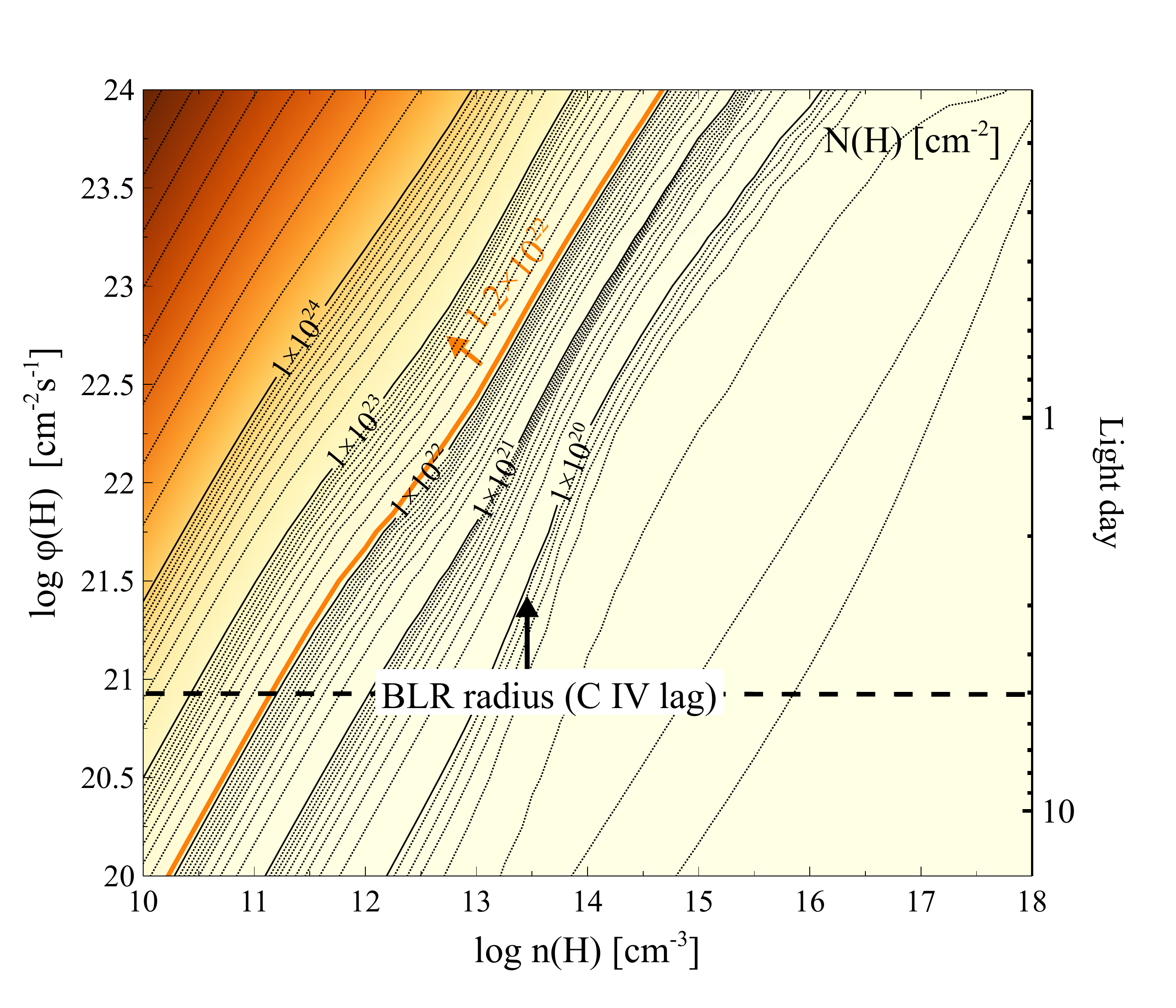}
 \caption{The contours show total hydrogen
 column density of the
 equatorial obscurer as 
 a function of the flux of ionizing photons and
 the hydrogen density.
 The orange line indicates
 the LOS obscurer’s column
 density (D19a) and the
 dashed black line shows the
 location of the BLR
 based on the observed \civ \ 
 lag.}\label{f1}
\end{figure*}

With the assumptions above, we computed
two-di\-men\-sion\-al grids
of 
photoionization models,
similar to those of
\citet{Korista97}. 
Each grid consists of a
range of total hydrogen 
density,
10$^{10}$ cm$^{-3}<n(\rm H)
<10^{18}$ cm$^{-3}$ , 
and a range of incident
ionizing photon flux,
10$^{20}$ s$^{-1}$
cm$^{-2} <\phi$(H)
$<10^{24}$ s$^{-1}$
cm$^{-2}$. 
The right vertical axis on
all plots (Figures \ref{f1} to \ref{f3}) shows the distance
from 
the incident ionizing
continuum source in light
days. The flux of ionizing
photons
$\phi$(H), the total
ionizing photon luminosity
$Q(\rm H)$, and the distance
in light days are related
by:

\begin{equation}
\phi(\rm H) = \frac{Q({\rm H})}{4 \pi r^2 }.
\end{equation}
For the SED of \cite{Mehd15}
and the observed luminosity
of
L (1-1000Ryd)=$2\times 10^{44} \rm \ erg
\rm\ s^{-1}, \rm \ the\  Q(\rm H)=
1.81\times 10^{54} \rm s^{-1}$ .

The STORM campaign reports
observed lags between 2 and 9
light days for various
strong emission lines
\citep[table 4]{DeRosa15}.
In Figure\ref{f1} , we
show contours of the
predicted obscurer’s
column density.  As
mentioned earlier, we
maintain a constant optical
depth of 8 at an energy of 4
Rydbergs, the lower limit to
have a holiday (D19b, figure
4).  

Next, we combine these
predictions with the
observations to derive the
properties of the equatorial
obscurer.

Before going on, we establish a nomenclature for the different
components that we discuss in this paper. For the case of UV lines,
\cite{Goad16} and many other previous work report the total  ``time-averaged broad emission line (BEL) EWs''.
We refer to this as the ``total'' emission. 
Subsequent work by \cite{Kriss19} model this total emission as the combination of three components: a ``broad'' component, a ``medium
broad'' component, and a ``very broad''
component. The sum of the two first components (broad
and medium broad) dominates in the line core, and
we refer to them as the BLR/core. 
For \civ\  line, these components have FWHMs of 3366$\pm 15$ and 8345$\pm 20$ km/s, with an average of $\sim 5000$ km/s.
Our calculations
in Section 3
suggest that the very broad component \citep[FWHM=16367$\pm 18$ km/s,
][]{Kriss19} forms in
the equatorial obscurer.  
For reference, table 1 of \cite{Kriss19} report
that the very broad component of \civ\  comprises almost 47\% of the total emission.

For Fe K$\alpha$, \cite{Cappi16}
report the presence of a time-steady ``narrow''
component with an upper-limit of 2340 km/s on the
line
width, or to be specific, FWHM $\leq 5500$ km/s.
This is very similar to the BLR component of \civ\ \citep[broad plus medium broad,
][]{Kriss19}.
Assuming the line is broadened by orbital motions, and adopting the
BH mass quoted by \cite{Cappi16}, they argue that
this component forms a few
light days away from the central source (0.006
pc), consistent with the lag observed for \civ. 
We refer to this component as the ``BLR''
Fe K$\alpha$ emission.   
The S/N ratios of the X-ray spectra do not permit a definitive detection of the
very broad component modeled in the \hst data,
although \cite{Cappi16} note that there appears to
be a broad, redshifted component underlying the Fe
K$\alpha$ profile.

\section{Wind properties
from the Observations}
The equatorial obscurer has
a higher column density than
the LOS 
obscurer since it is closer
to the accretion disk, the 
site where the wind is
launched. The orange line in
Figure~\ref{f1} shows 
the column density of the
LOS obscurer,\  $N(\rm
H)=1.2\times 10^{22} \rm
cm^{-2}$ \citep{Kaastra14}.
The orange arrow shows the
direction of possible higher
column density obscurers.  

The horizontal dashed black
line indicates the location
of the BLR 
adopting the \civ \  lag
reported by \cite{DeRosa15}. To  
ensure that the base of the
wind is located between the
central SMBH and the BLR,
we must choose an obscurer
with a smaller distance
(higher flux of ionizing photons) from the
continuum source, than that
for the BLR, the region
suggested by the black
arrow.  

As Figure \ref{f1} shows, lines with constant column density are almost parallel for $N(\rm H)> 10^{21} \rm cm^{-2}$, and their values
increase toward the upper left corner, closer to the source. These lines also
represent a nearly constant ionization parameter, which increases toward the upper left corner. 

The properties of the equatorial obscurer are
constrained by observations. The equatorial obscurer is
a source of emission 
itself since energy is conserved,
and it must re-radiate the energy
that is 
absorbed. If the equatorial
obscurer emission is strong
enough, then it produces
a second emission-line region
between the original BLR and the
source. 
Since re-emission by the obscurer is not evident in the observations, we
must find a model of the obscurer
which not only 
explains the holiday, but also
does not dominate the strong lines
seen by \hst and \xmm.  
To do this, we considered the
total observed equivalent widths (EWs)
of strong emission lines from the
STORM data \citep{Goad16,Pei17} 
and the total luminosity of Fe K$\alpha$
observed by \xmm \citep{Mehd15}.

In general, an obscuring cloud may
cover only a small fraction of the
continuum source, 
as in the leaky LOS obscurer shown
in figure 6 of D19a, or it can
fully cover the 
continuum source (CF=100\% in their
figure). Here we assume that the
equatorial obscurer fully
covers the central object along the LOS of the BLR, which is the preferred situation explained by D19b.

We wish to directly compare our predictions with the observations. 
We report all lines as EW relative to the continuum at 1215\AA \  so that ratios of EWs are the same as ratios of intensities.  

The EW is proportional to the ratio of a line luminosity to the continuum.  
We assume that the continuum is isotropic and that \hst\  had an unextinguished 
view of it.  The continuum luminosity is not affected by the equatorial obscurer's CF. The luminosities 
of lines emitted by the equatorial obscurer are linearly proportional to the equatorial 
global CF, the fraction of 4 $\pi$ steradian covered by the obscurer.  
The equatorial obscurer covering factor is not known but must be at least 50\%\  if 
it is to shield the BLR.  We report EWs for full coverage with the understanding that the actual EW of the obscurer is: 

\begin{equation}
\rm EW(\rm obscurer) = \frac{\Omega}{4 \pi} \times \rm EW(pred) \sim 50\% \ \rm EW(\rm pred).
\end{equation}
On the other hand, the equatorial obscurer is not a dominant contributor to the emission lines. 
As a first step in the modeling, we set a limit to the amount of emission from the 
obscurer is less than half of the total emission. To choose this value, we were motivated by the ratio of the flux of very broad \civ\  to the flux of total observed \civ,
47\%, as measured by \cite{Kriss19}:
\begin{equation}
 \rm EW(\rm obscurer) \leq 50\% \  \rm EW(observed).
\end{equation}

\begin{figure*}
\centering
%\begin{minipage}{3 in}
\includegraphics [width=\textwidth]{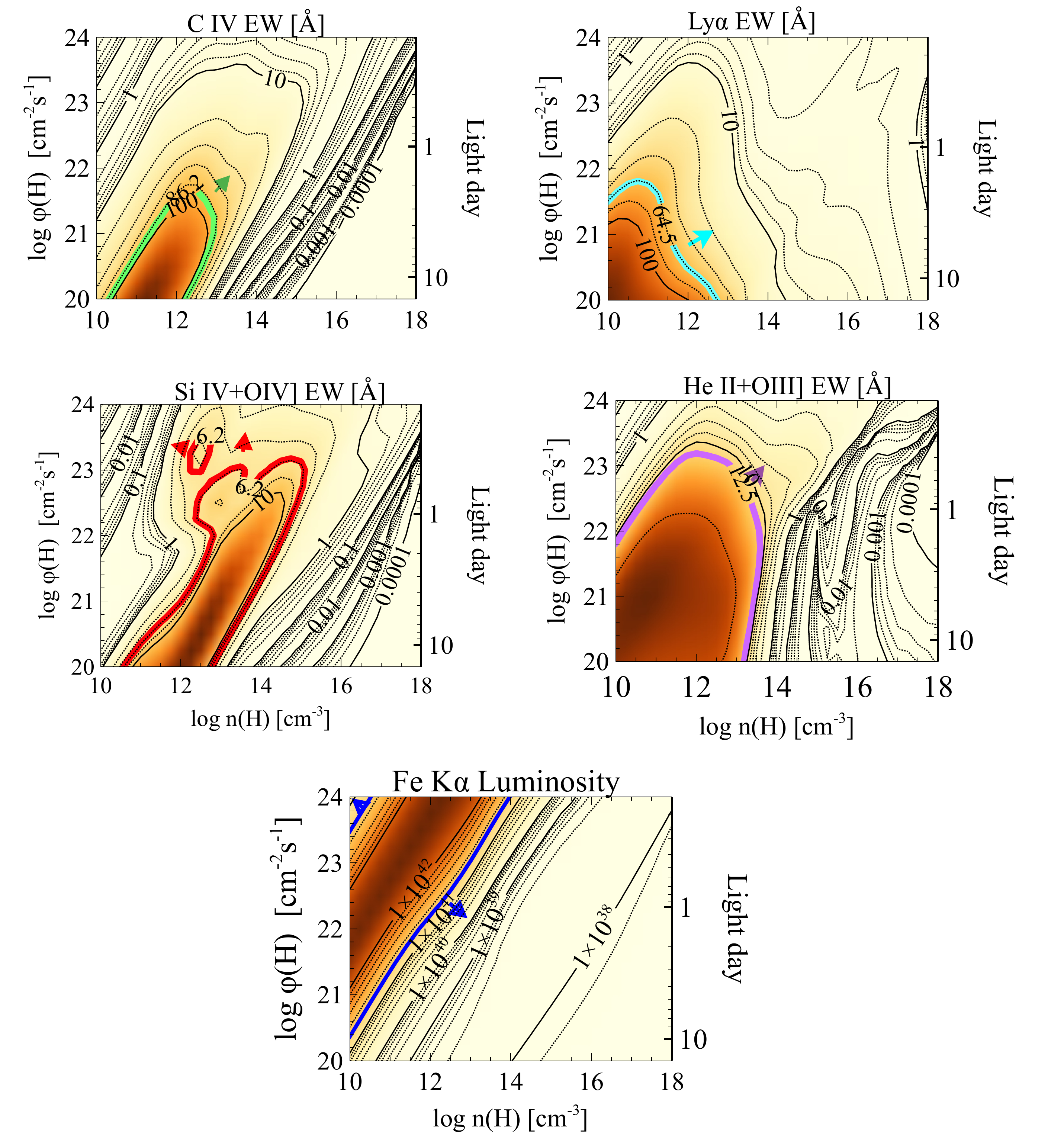}
%\end{minipage}
\caption{Upper four panels show the predicted EW of strong lines emitted by the equatorial 
obscurer as the contours. The colored lines indicate the \hst observed value and arrows
show the direction in which the equatorial obscurer must be chosen in order for its emission to not dominate the \hst
emission lines. All the EWs are
relative to the 1215 \AA \
continuum. The lowest panel shows
the predicted luminosity of Fe
K$\alpha$  as the contours and the blue lines show
the \xmm observed values. }\label{f2}
\end{figure*}

Based on equations 2 and 3, the two factors of 50\% cancel:
\begin{equation}
 \rm EW(\rm pred) \leq \rm EW(observed),
\end{equation}
which means any model of the 
equatorial obscurer that produces lines with EW less than the observed total values are allowed. 
We map the obscurer's predicted emission lines in Figure~\ref{f2}. We also include 
the observed values as colored lines in each panel. The arrows show the physical conditions where the obscurer will \textit{not}  dominate the emission line fluxes of observed \hst spectrum. 

The lowest panel of Figure~\ref{f2} shows the predicted luminosity of Fe K$\alpha$ for full coverage. When the obscurer is highly ionized, 
Fe K$\alpha$ is strong (dark orange). It becomes weaker in the extreme upper left 
corner where the obscurer is fully ionized. In this regime, there are few bound 
electrons and there is no iron emission line or edge. The observed 
time-averaged value of its luminosity for the 2013 campaign is $(2.0 \pm 0.3)\times 10^{41}$ erg/s \citep{Mehd15} and is indicated by the blue lines in Figure\ref{f2}.

Satisfying the constraints from Equations 2 \& 3
guarantees that the obscurer does
not produce strong emission lines.
For the rest of the
modeling, we assume this holds for all lines except \heii\ and broad
Fe K$\alpha$. 
As discussed below, the lag profiles measured
by \cite{Horn19} show that \heii\  forms very
close to the central source.
We assume that all of the UV \heii\  comes
from the obscurer.
The Fe K$\alpha$ profile discussed below is
consistent with half of the line forming in
the BLR with a broad
base forming in the obscurer.

\begin{figure*}
\centering
%\begin{minipage}{4 in}
\includegraphics [width=\textwidth]{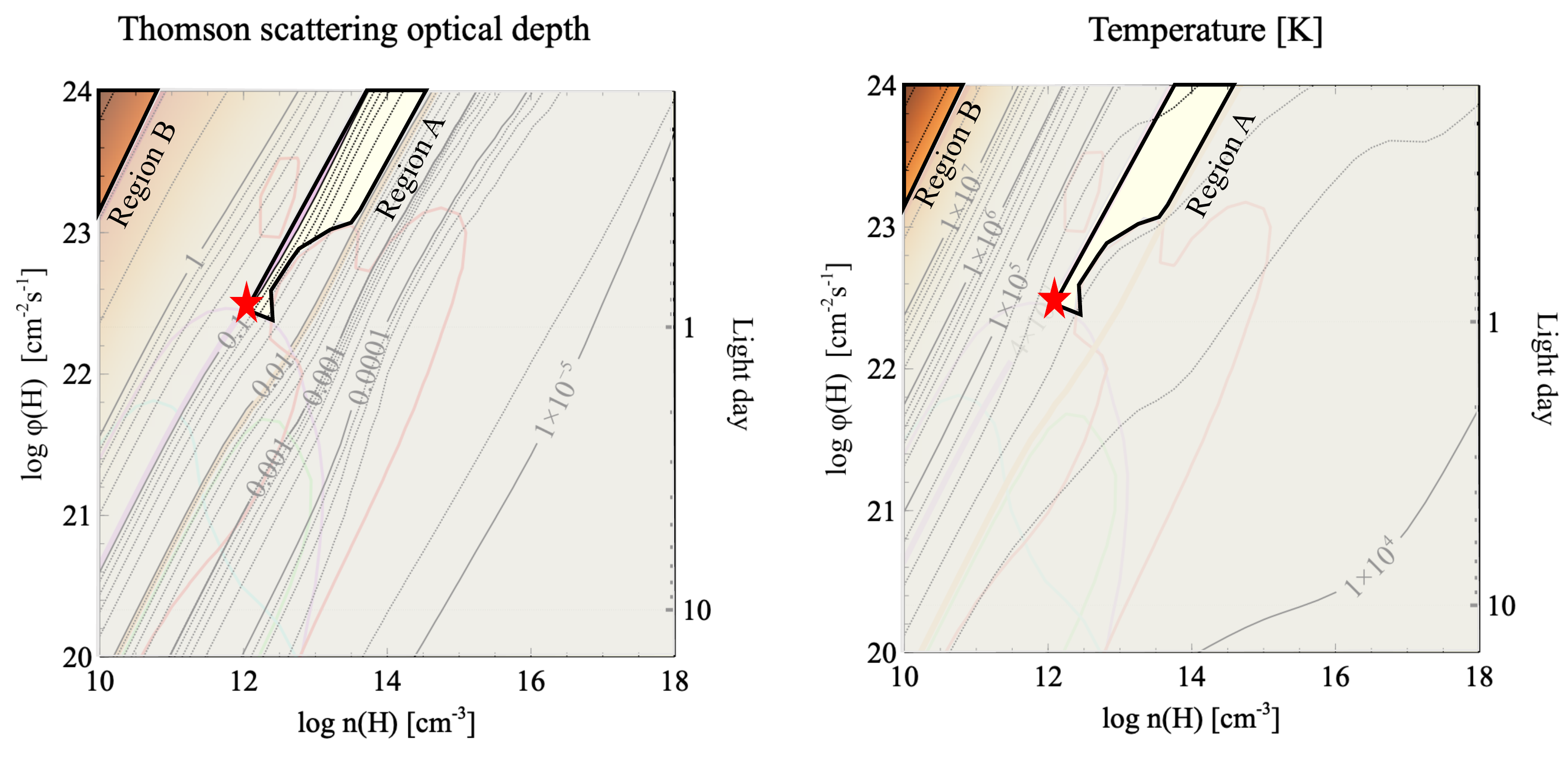}
%\end{minipage}
\caption{The left panel maps the Thomson scattering optical depth and the right panel maps the temperature
 of the obscurer. A and B are two
 regions with allowed properties of the equatorial obscurer. The red star
 indicates our preferred
 model, which is the most consistent with all observational constraints}.\label{f3}
\end{figure*}

Figure~\ref{f3} shows the regions which
satisfy all the constraints inferred from
Figures~\ref{f1}\&\ref{f2}. All the forbidden areas are colored in gray.
The right panel 
shows the variation of the temperature 
as a function of both the flux and the density. The temperature
increases as the distance to the central source decreases. The left panel maps the Thomson
scattering optical depth as a function of flux and density.  
Gas in the upper left corner of the plot has a significant 
column density and Thomson scattering optical depth. 
Note that the soft X-ray observations constrain the ionization parameter but not the density or distance
from the center
so any location along the line is allowed. In
both panels, all the constraints from
Figure~\ref{f2} are shown as faint colored lines, in
order to show how we recognize the forbidden region.  

As shown in both panels, there are two
possible regions for the obscurer's properties:

Region A: $r~<$1 light days,
$10^{12}~\rm cm^{-3}~<n(\rm
H)~<~10^{14}~\rm cm^{-3}$, 
$\phi(\rm H)>~10^{22.4}~\rm s^{-1}\rm cm^{-2},
10^{4.6} K<T<10^{4.8} K$, 
and $1.2\times 10^{22}\rm cm^{-2}\leq N(\rm H)< 2 \times 10^{23} \rm cm^{-2}$. 
The low-density bound of the region is set by the luminosity of Fe K$\alpha$, the lower bound by He II, and the 
high-density bound by LOS column density.  
It has a Thomson scattering optical
depth between 0.01 and 0.1. 

Region B: $r<$0.4 light days,
with $n(\rm H)< 10^ {11} \rm cm^{-3}$, and
$\phi(\rm H)> 10^{23} \rm s^{-1}\rm cm^{-2}$,
$\rm T \ge 3\times10^{7}, \rm$ and
$\rm N(\rm H)\geq 10^{24} \rm cm^{-2}$. It has a very high ionization parameter
and is Compton thick (Figure \ref{f2}). 
The lower limit to this region is set by the Fe
K$\alpha$ emission. The Thomson
scattering optical depth is $\tau_e \geq$4. 

We prefer region A since it produces
significant very broad \heii\ and Fe K$\alpha$
emission, but produces other UV lines with EWs less than half the observed values.
The \heii\ velocity-delay map sets a $\le 5$ day limit to the lag \citep{Horn19}. 
This is consistent with almost all of the
observed \heii \ being  produced in the
equatorial obscurer. As with the UV lines, we
assume that half of the Fe K$\alpha$ forms in the obscurer, 
with the other half in the BLR.
Below we show  that this is also suggested by the Fe K$\alpha$ line profile, in which half of the line EW forms in the BLR and the rest is
a strong broad component that forms in the equatorial obscurer.
This might be the very broad Fe K$\alpha$ component mentioned by  \cite{Cappi16} and is produced in the obscurer. 

Region B is not of interest for our model of the wind
since the EWs of the broad UV lines produced by any winds
chosen from this region are almost 1\% of the total observed
values. Moreover, a wind chosen from region B will be very
close to the central source and will emit lines much broader than
what was observed.    

The parameters for our final preferred model,
$\phi(\rm H)\approx 10^{22.5}\rm s^{-1}\rm cm^{-2}, 
n(\rm H)\approx 10^{12} \rm cm^{-3}, T\approx 5 \times 10^{4}\rm K$, and $\tau_e \approx 0.1$
are shown with a star in Figure~\ref{f3}. A wind with these parameters is our favorite model in region A, since it has a major contribution to the \heii\  and Fe K$\alpha$ emissions. Any other wind selected from region A will emit lower values of the mentioned lines.
These conditions place the wind/equatorial obscurer at about one light day from
the central source.  Please note that although the
mentioned hydrogen density seems to correspond to the
changing look portion of figure 4 of D19b, since the current
paper has adopted a different $\phi(\rm H)$ for the
equatorial obscurer, the ionization parameter is nearly the
same as case 2 in D19b.   This means an obscurer with
mentioned
$\phi(\rm H)$ and $n(\rm H)$ belongs to
the case 2 discussed in D19b and reproduces the holiday. This was expected since by keeping the optical depth constant, we made sure that all of the models in this paper belong to case 2 of d19b.

\begin{figure*}
\centering
%\begin{minipage}{3 in}
\includegraphics [width=\textwidth]{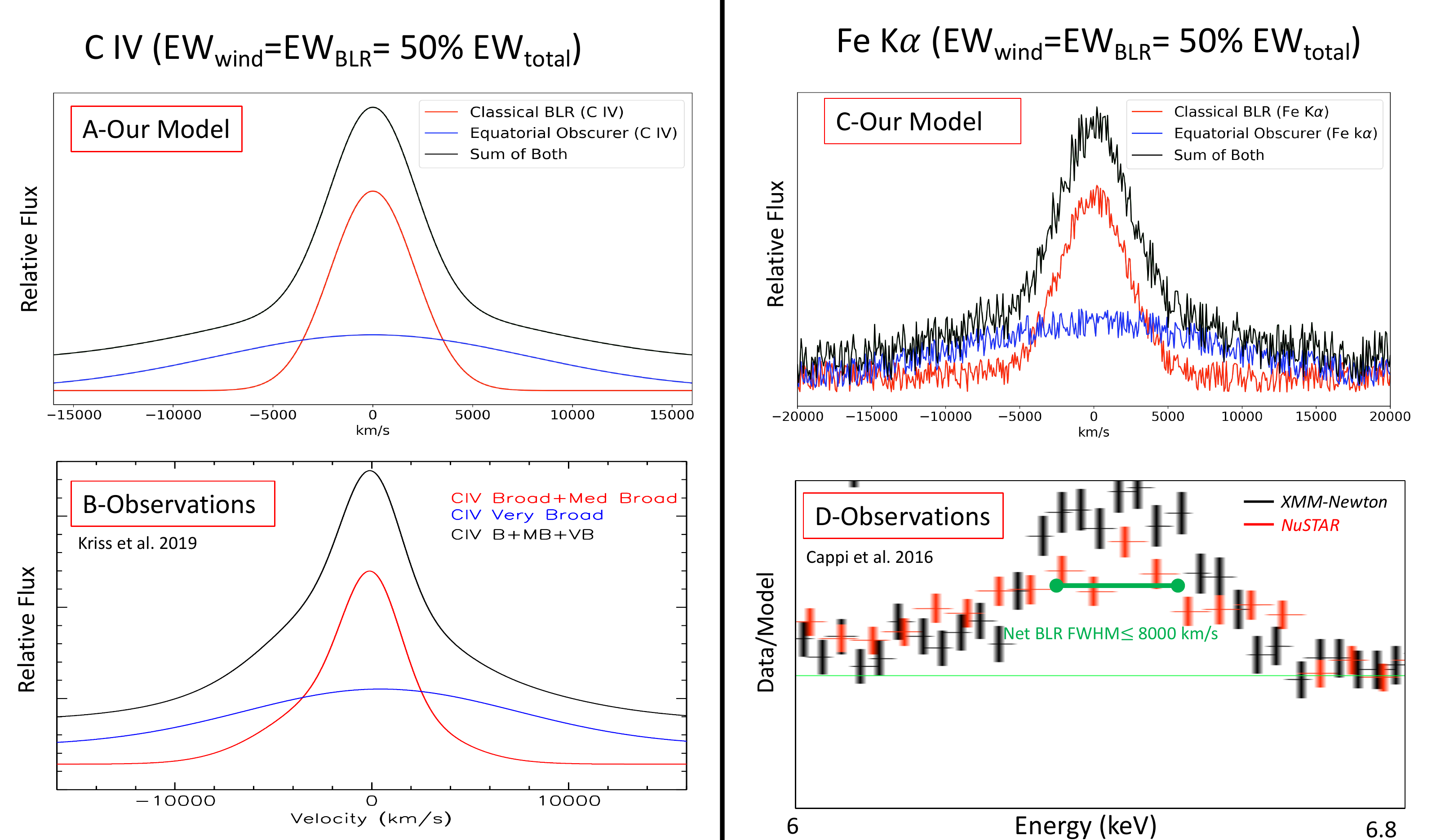}
%\end{minipage}
\caption{This Figure compares our
model with the observations from \hst,
\xmm, and \nustar. Panels A \& B show the
case for \civ, for which the obscurer
produces a very broad component (panel A,
blue) with an EW of half of that
produced by the BLR (panel A, red).
Panels C \& D show the case for Fe
K$\alpha$, for which the obscurer
produces a very broad component (panel C,
blue) with an EW equal to that
produced by the BLR (panel C, red). 
It is plausible that a broad base similar to \civ\  is present,
although the S/N is not high enough to say for sure.
In
both cases our predictions are
very similar to the observations,
suggesting that the disk wind could be
responsible for the observed very broad emission line components.}\label{fka}
\end{figure*}

Figure \ref{fka} compares our predictions
for the \civ \ and Fe K$\alpha$ line profiles with the 
observations.  
To illustrate our preferred model (panels A and C), 
we adopt 
a SMBH mass of  $\rm M =(5.2\pm 0.2) \times 10^7 \rm M_{\odot}$ \citep{Bentz15}.
Assuming Keplerian motion and the equations given in the
first paragraph in section $5.1$ of \cite{Cappi16}, the lines
produced by the equatorial obscurer have a FWHM of
18500$\pm 3500$ km/s. The more recent  BH mass estimations
are about  50\% larger than our adopted
value \citep{Horn19}. This represents the
uncertainty in the BH mass measurements and causes 20\%
uncertainty on the FWHM  of our model, since the
line width estimation depends on the BH mass. We adopt the mass determined by \citep{Bentz15}, to be
consistent with \cite{Kriss19}.

Figure \ref{fka}  panels A (theory) and B (\hst observations) show the case
for \civ, in which we are using
arbitrary vertical offsets in flux,
simply for illustrative purposes. To produce
panel A, we assume that the equatorial
obscurer is emitting \civ \ with an EW 
half that observed and with FWHM=18500 km/s (blue line), while the BLR emits the flux with FWHM=5000km/s (red line, \citealt{Kriss19}). 
Panel B is the best fit model to the \hst STORM observations \citep{Kriss19}. 
Those panels suggest
that  the
equatorial obscurer could well be responsible for the very
broad component.

Figure \ref{fka}  panels C (theory) and D (\nustar and \xmm observations, 2013 Jul 11-12, Jul 23-24, and Dec 20-21) show the same thing for the
Fe K$\alpha$ line, but this time we assume that
the obscurer produces the emission line with
an EW equal to that observed and a FWHM=18500$\pm 3500$
km/s (blue line), while the BLR emits Fe K$\alpha$
with FWHM=5500 km/s (red line, \citealt{Cappi16}).
Panel D shows the observations of
\cite{Cappi16} in which the vertical axis indicates the data as the ratio to a single power-law continuum
model fitted to the \xmm\  (black) and \nustar\ (red) observations.  The green horizontal
line shows the net FWHM which is
calculated by adding the widths of two Gaussian functions with the same central wavelength position in quadrature (the core
corresponding to the observed broad Fe
K$\alpha$ FWHM=5500 km/s  and
the \xmm resolution with $\rm
dE/E=1/50$, so FWHM=6000 km/s). This
results in a net BLR FWHM$\leq$8000 km/s,
consistent with the BLR core observed by \hst
and suggests that the core of the observed Fe K$\alpha$ profile is
in good agreement with the our model.  Comparing
panels C  and D, which
are equally scaled,
shows that the very broad emission from the
obscurer might easily hide under the total
emission and be just seen as a very broad
continuum. This very broad base may be observable in
Panel D at $\pm 7000 $ km s$^{-1}$.

The total observed Fe K$\alpha$ profile (panel C) is similar to the \civ\  seen by \storm,
although this is not a strong statement due to the S/N ratio in the X-ray data.
Indeed, the total \civ\  is consistent with the ``narrow'' Fe K$\alpha$ discussed in \cite{Cappi16}. Motivated by this similarity, we propose that this
line also includes the classical BLR emission and a very broad component originated from the wind, hidden in the noise. This scenario is a testable hypothesis for our model and can be the subject of future observations with Chandra / HETG.

\section{Discussion and summary}

Here, we have used \hst and \xmm observational
constraints to derive a model of the equatorial
obscurer.  We have shown that the equatorial
obscurer, which modifies the SED to produce the
emission-line holiday, is itself a significant
source of line emission, solving several
long-standing problems in emission-line physics.
The model predicts that lines should have a core
formed in the classical BLR and strong broad wings,
a profile consistent with the line deconvolution
presented in \cite{Kriss19}, and that much of the UV \heii\  and X-ray Fe K$\alpha$ can originate in
the equatorial obscurer.   Finally, we found that
the obscurer has a modest optical depth to electron
scattering and so adds reflection and scattering to the physics of the line-continuum transfer function
and emission-line profiles. This is a unified model
of the disk wind in which the remarkable responses
of the emission lines in NGC 5548 are explained and
the properties of the unobservable part of the wind
are derived.

\begin{figure*}
\centering
\begin{minipage}{4 in}
\includegraphics [width=\textwidth]{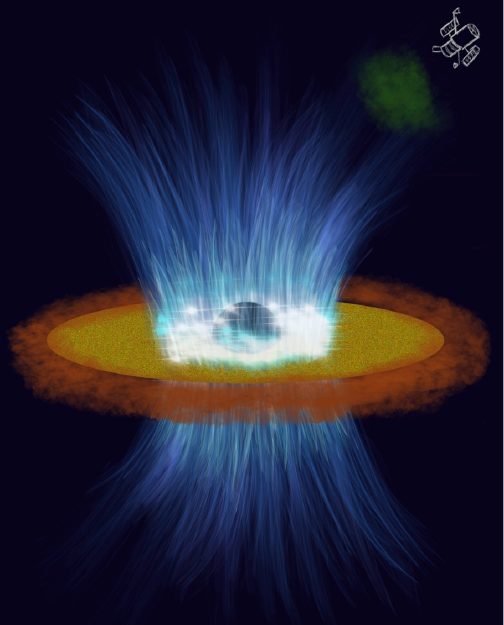}
\end{minipage}
 \caption{Cartoon of the disk wind in \ngc \  (not
 to scale). The disk wind (blue) 
 surrounds the central black hole
 and extends to the line of sight to 
 HST in the upper right corner. The BLR is shown as
 the orange cloud around the 
 disk. The green cloud at the upper right shows
 the absorbing cloud discussed
 in D19a. The bright region in the
 lower part of the wind indicates
 that the wind is a major
 contributor to the very broad
 components of the observed emission lines
 }. \label{f4}
\end{figure*}

Figure~\ref{f4} shows a cartoon of our derived
geometry. This Figure is consistent with figure~1 of D19b, however, here we also consider the emission from the wind.  
The very bright area, the base of the wind,
indicates this emission from the equatorial obscurer. Variations in this part of the wind 
produce the emission-line holiday (D19b). 

This model is also consistent with the
\cite{Sim10}
 Monte
Carlo radiative transfer predictions of the
X-ray spectra of a line-driven AGN disc
wind. They argued that a disk wind can
easily produce a significant strong, broad
Fe K$\alpha$
component which has a complex line
profile. Based on their simulations, the wind's effects on
reflecting or reprocessing
radiation is at least
as important as the  wind's effects on the
absorption signatures. Their model was later followed by \cite{tatum12}, in which a Compton-thick disk wind is responsible for all moderately broad Fe K emission components observed in a sample of AGNs. Their disk wind is not located in the LOS to the source and still affects the observed X-ray spectrum.

The electron scattering optical depth could be
larger than estimated here, $\tau_e \sim 0.1$.
Our derived parameters are highly approximate
suggestions of the properties
of the equatorial obscurer.  We choose
the smallest Lyman continuum optical
depth (and H$^{0}$ column density)
obscurer that is consistent with D19b.
Other solutions with similar atomic
column density but greater thickness
are possible. 
They would have larger ionized column
density and electron scattering optical
depth.
The Thomson optical depths reported in Figure
\ref{f3} are normal to the slab.  
A ray passing into the slab at an angle $\theta$
will see an optical depth of $\tau_0 / \cos
\theta$.
For isotropic illumination the mean
optical depth is $\sqrt2$ larger than
the normal.

A region with a significant electron
scattering optical depth 
and warm temperature, $T\approx 5
\times 10^{4}\rm K$, would solve several outstanding problems,
which we summarize next.

It could be a part of the Compton reflector
and so constitutes a translucent mirror in the
inner regions.  
Scattering off warm gas will help producing smooth line profiles \citep{arav99},
a long-standing mystery in the geometry of the BLR.
Gas with these properties also produces bremsstrahlung emission with a temperature 
similar to that deduced by  \citet{Ant88} and so
could provide the location of the non-disk emission.
The obscurer modeled here is not a significant source of bremsstrahlung emission, however.

A model with an electron scattering
optical depth $\geq 0.5$ could provide an
obscuration required for explaining the
velocity-delay maps of \cite{Horn19}. 
They show that the emission from the far
side of the BLR is much fainter than expected
with isotropic emission from the central source
and no obscuration. 
If the base of the wind is transparent we will
observe both the near and far sides of the BLR.
This indicates that there must be an obscuring cloud between
the BLR and the source, acting like a mirror.

D19b proposed that the disk wind can be
transparent or translucent. This hypothesis is
compatible with figure 4 of \cite{Gius19}, in which
\ngc\ is on the border of having a
line-driven disk wind or a failed wind. This means that
small changes in the disk luminosity/ mass-loss rate
will affect the state of the wind. The reason is
that decreasing the disk luminosity leads to a
reduction in the mass flux density of
the wind, making it over-ionized \citep{Prog04}.
A transparent wind has little
effect on the SED
and no spectral holidays occur, 
while holidays  occur  when the wind is translucent.
In this state, the equatorial obscurer
absorbs a great deal of the XUV / X-ray part of
the SED which
must be reemitted in other spectral regions. %D19b also show that the equatorial obscurer
%only needs an 8\% change in its hydrogen density to
%reproduce the holiday discussed in \cite {Goad16}.
%The changes in EW reported in \cite {Goad16} are
%time-averaged values, so their amplitude is weakened
%by the time-delayed response across the geometry of
%the BLR. Thus, the effect of the equatorial obscurer
%is likely to be more dramatic and the variations
%of its hydrogen density might be more than 8\%. These variations will transfer to the LOS obscurer
%and result in the
%absorption-line holiday, due
%to changes in the CF (D19a).

In this paper, we introduced a new approach to
derive the wind's properties. This will have
important implications for future studies of AGN
outflows and feedback.
We used  observations to discover
the behavior of a part of the wind the can never be
directly observed.
Our models of the wind will be expanded to better approximate the hydrodynamics of the wind. Deriving these ``next generation'' hydrodynamical / microphysical models
and comparing them with the observations  will be the subject
of our future study.

\acknowledgments
Support for {\it HST} program number GO-13330 was provided by NASA through a grant from 
the Space Telescope Science Institute, which is operated by the Association of Universities
for Research in Astronomy, Inc., under NASA contract NAS5-26555. We thank NSF (1816537, 1910687), NASA (17-ATP17-0141, 19-ATP19-0188), and STScI (HST-AR-15018, HST-AR-14556). 
MC acknowledges support from NASA through STScI grant HST-AR-14556.001-A  and NASA grant 19-ATP19-0188, and also support from National Science Foundation through grant AST-1910687. 
M.D.\ and G.F.\ and F. G.\  acknowledge support from the NSF (AST-1816537), NASA (ATP 17-0141),
and STScI (HST-AR-13914, HST-AR-15018), and the Huffaker Scholarship.
M.M. is supported by the Netherlands Organization for Scientific 
Research (NWO) through the Innovational Research Incentives Scheme Vidi grant 639.042.525.
J.M.G. gratefully acknowledges support from NASA under the ADAP award 80NSSC17K0126. MV gratefully acknowledges support from the Independent Research Fund Denmark via grant number DFF 8021-00130.

 \end{document}